\documentstyle[preprint,aps,eqsecnum,epsf,floats]{revtex}

\input amssym.def
\input amssym.tex

\tighten 

 \input BoxedEPS
\SetRokickiEPSFSpecial  
 \HideDisplacementBoxes

\def\half{{\textstyle{1\over2}}}

\def\beq{\begin{equation}}
\def\eeq{\begin{equation}}

\def\pmb#1{\setbox0=\hbox{$#1$}%
     \kern-.025em\copy0\kern-\wd0
        \kern.05em\copy0\kern-\wd0
           \kern-.025em\raise.0433em\box0}

\skip\footins = 14pt 
\begin{document}
\nonfrenchspacing
\flushbottom
\title{Dynamical Poincar\'{e} Symmetry\\ Realized by Field-dependent
Diffeomorphisms}
\bigskip

\author{R.~JACKIW\cite{RJauth}}
\address{{\it Center for Theoretical Physics,  Massachusetts Institute of Technology}
\centerline{\it Cambridge, MA ~02139--4307, USA}}
\author{ A.P.~POLYCHRONAKOS}
\address{{\it Theoretical Physics Department, Uppsala University}
\centerline{\it S-75108 Uppsala, Sweden}}

\maketitle
\vspace{.2in}

\centerline{Faddeev {\it Festschrift\/}, Steklov Mathematical Institute Proceedings}

\vspace{.2in}
\begin{center}

\centerline{\bf Abstract}
\medskip

\parbox{5in}{\small\baselineskip=12pt We present several Galileo invariant Lagrangians,
which are invariant against Poincar\'{e} transformations defined in one higher (spatial)
dimension.  Thus these models, which arise in a variety of physical situations, provide a
representation for a dynamical (hidden) Poincar\'{e} symmetry.  The action of this
symmetry transformation on the dynamical variables is nonlinear, and in one case
involves a peculiar field-dependent diffeomorphism.  Some of our models are completely
integrable, and we exhibit explicit solutions.}
\end{center}



\thispagestyle{empty}


\section{Introduction}

Our colleague and friend Ludwig Faddeev has brought mathematics to physics and physics
to mathematics.  Here we mention only his work on integrable systems and on group
theory, because these two themes also inform our essay, which is dedicated to him on the
occasion of a significant birthday.

We shall be concerned with Lagrangians containing fields that move in time on a
$d$-dimensional space.  The models arise from diverse physical systems: continuum
description of free particle motion, isentropic fluid mechanics with irrotational
velocity field, hydrodynamical description of quantum mechanics, free motion of
membranes as well as higher-dimensional ``$d$-branes".  Our models are Galileo
invariant on their $(d,1)$ space-time.  However, they possess a further hidden symmetry: in
terms of the dynamical canonical variables with which these theories are formed one can
construct quantities whose algebra, determined by canonical Poisson brackets, reproduces a
$(d+1,1)$ Poincar\'{e} algebra. Moreover, one finds that the $(d+1,1)$ Poincar\'{e} group is a
symmetry of the $(d,1)$ Galileo invariant theory, with various $(d+1,1)$ Poincar\'{e}
transformations acting non-linearly on the available $(d,1)$ coordinates.  Indeed some
of these coordinate transformations make use of peculiar diffeomorphisms that involve
the fields themselves.  In summary we find that our models give a non-linear
representation for a dynamical (hidden) Poincar\'{e} group.

Another interesting feature is that several of our models are completely integrable.
This is seen by explicitly integrating the Euler-Lagrangian equations for some models,
while for others -- in one spatial dimension -- by identifying an infinite number of
conservation laws, and replacing the nonlinear equations  by linear ones,
whose solutions are explicitly given.

\section {The Models}
The Lagrangians that we study involve two canonically conjugate fields, $\theta$ and
$\rho$. When presented in first-order form, they read
\begin{eqnarray}
L &=& \int d^dr\, \Big( \theta \dot{\rho}-\half \rho \bbox{\nabla} \theta
\cdot \bbox{\nabla}\theta - V (\rho) \Big)
\label{eq:2.1}\\
L_o&=&\int d^dr\, \Big(\theta \dot{\rho}-\half \rho \bbox{\nabla} \theta \cdot
\bbox{\nabla}
\theta \Big)
\label{eq:2.2}
\end{eqnarray}
Fields depend on time and space, $(t,\bbox{r})$; the over-dot denotes differentiation
with respect to the temporal argument; the gradient is with respect the spatial
arguments.  $L_0$ is the ``free" Lagrangian (even though it is not quadratic), while $L$
includes an interaction potential $V(\rho)$, which we take to be $\theta$-independent.

The canonical structure implies the Poisson bracket
\begin{equation}
\{ \rho(t,\bbox{r}), \theta(t,\bbox{r}^\prime)\} = \delta
(\bbox{r}-\bbox{r}^\prime)
\label{eq:2.3}
\end{equation}
and the Hamiltonians
\begin{eqnarray}
H &=& \int d^d r\, {\cal E} \phantom{_012} \qquad {\cal E} = \half \rho \bbox{\nabla}
\theta \cdot\bbox{\nabla}\theta +V(\rho)
\label{eq:2.4}\\
H_0 &=& \int d^d\, \bbox{r} {\cal E}_0 \phantom {_01}\qquad {\cal E}_0 = \half \rho
\bbox{\nabla}\theta \cdot\bbox{\nabla}\theta
\label{eq:2.5}
\end{eqnarray}
(Here we are following the simplectic method for Lagrangians that are linear in the
time derivative -- this approach was advocated by Faddeev and one of the present
authors (RJ).\cite{bib:1}) Evidently the Euler-Lagrange equations read
\begin{eqnarray}
\dot{\rho} &=& -\bbox{\nabla}\cdot(\rho\bbox{\nabla}\theta)
\label{eq:2.6}\\
\dot{\theta} &=&-\half (\bbox{\nabla}\theta )^2 +f(\rho)
\label{eq:2.7}\\
f(\rho)&=&-{\delta \over \delta \rho}\int d^d r\, V(\rho)
\label{eq:2.8}
\end{eqnarray}
where the ``force" term $f$ is absent in the free case.

We now describe several contexts that lead to this system.
\subsection{Continuum Description of Free Particle Motion}

Begin with the free Lagrangian for a collection of particles, with equal masses scaled to
unity.
\begin{equation}
L_{\rm free~particle} = \half \sum_i v_i^2(t) \qquad(m=1)
\label{eq:2.9}
\end{equation}
Let the particle counting index $i$ become the continuous variable $\bbox {r}$, and
introduce the density $\rho(t,\bbox{r})$ and current $\bbox{j}(t,\bbox{r})$, with
\begin{equation}
\bbox{j}=\bf{v}\rho
\label{eq:2.10}
\end {equation}
Then $\half \sum\limits_i v _i^2$ becomes $\half \int d^d r\, v^2
\rho =\half \int d^d\, r j^2/ \rho$.
We wish to link the density and current by a continuity equation
\begin{equation}
\dot\rho+\bbox \nabla \cdot \bbox j = 0
\label{eq:2.11}
\end{equation}
which can be enforced with help of a Lagrange multiplier $\theta$.  We thus arrive at
the continuum Lagrangian
\begin{equation}
L_{\rm free~particle} \rightarrow L _0^\prime =\int d^dr\, \Big\{\half j^2/
\rho + \theta (\dot\rho + \bbox{\nabla} \cdot \bbox {j}) \Big\}
\label{eq:2.12}
\end{equation}
Since $\bbox j$ does not participate in the canonical $1$-form, $\int d^d r\, \theta
\dot\rho\, d t = \int d^dr\, \theta\, d \rho$, we can eliminate $\bbox j$ by solving the
constraint equation that follows when $\bbox j$ is varied in $L_0^\prime$ \cite{bib:1}.
\begin{equation}
\bbox j = \rho \bbox{\nabla} \theta
\label{eq:2.13}
\end{equation}
Comparing (\ref{eq:2.10})  with (\ref{eq:2.13})  shows that the velocity is given by
$\bbox{\nabla} \theta$;  hence it is irrotational, with $\theta$ playing the role of a velocity
potential.
\begin{equation}
\bbox {\nabla} \times {\bf{v}}=0 
\label{eq:2.14}
\end{equation}
\begin{equation}
\bf{v} = \bbox{\nabla} \theta
\label{eq:2.15}
\end{equation}
Substitution of (\ref{eq:2.13}) in (\ref{eq:2.12}) yields $L_0$ of
(\ref{eq:2.2}). 
\begin{equation}
L_0^\prime \rightarrow  L_0= \int d^dr\, \Big\{ \theta \dot\rho - \half \rho
\bbox{\nabla} \theta \cdot \bbox{\nabla} \theta \Big\}
\label{eq:2.16}
\end{equation}
Supplementing $L_0$ with a velocity-independent ($\theta$-independent) interaction term
produces\cite{bib:2}
\begin{equation}
L = L_0 +\int d^dr\, V(\rho)
\label{eq:2.17}
\end{equation}

\subsection{Fluid Mechanics}

Equations (\ref{eq:2.6}), (\ref{eq:2.7}) are recognized as the equations of
fluid mechanics when the velocity field $\bf{v}$ is irrotational  and the motion is
isentropic.\cite{bib:3}  
The equation of motion (\ref{eq:2.6}) is the continuity equation (\ref{eq:2.11}),
once (\ref{eq:2.13}) is taken into account.  Moreover, that equation (\ref{eq:2.7}) for the
velocity potential produces the Euler equation of fluid mechanics

\begin{equation}
\bf{\dot{v}} + \bf{v}\cdot\bbox{\nabla}\bf{v} = - {1\over\rho}
\bbox{\nabla} (\rm pressure)
\label{eq:2.18}
\end{equation}
is established by taking the gradient of (\ref{eq:2.7}), and using (\ref{eq:2.15}).
\begin{equation}
{\bf{\dot{v}} + \bf{v}\cdot\bbox{\nabla}\bf{v}} = f^\prime (\rho)
\bbox{\nabla} \rho
\label{eq:2.19}
\end{equation}
The right-hand sides of (\ref{eq:2.18}) and (\ref{eq:2.19}) coincide for isentropic motion,
where pressure is a function only of $\rho$  [Kelvin's theorem].  [In that case ${\partial V
(\rho)
\over
\partial \rho} = - f(\rho)$ is the enthalpy $w$, and $(\rho 
{\partial^2V(\rho)\over\partial \rho^2})^\half$ is the sound speed  $u$.]

\subsection{Hydrodynamical Description of Quantum Mechanics}

For another derivation, we consider the Lagrangian for the Schr\"{o}dinger equation.
\begin{equation}
L_{\rm Schr\ddot{o}dinger} = \int d^dr\, \Big\{ i \psi^\ast\dot{\psi}-\half
(\bbox{\nabla} \psi)^\ast \cdot(\bbox{\nabla} \psi)- \overline{V} (\psi^\ast \psi)\Big\}
\label{eq:2.20}
\end{equation}
The first two terms correspond to the free, linear quantum mechanical equation, while $
 \overline V$ allows for possible non-linearity.  Upon setting in (\ref{eq:2.20})
\begin{equation}
\psi = \rho^\half e^{i\theta}
\label{eq:2.21}
\end{equation}
we get \cite{bib:4}
\begin{equation}
L_{\rm Schr\ddot{o}dinger} \rightarrow L = \int d^dr\, \Big\{ \theta \dot{\rho} - \half
\rho \bbox{\nabla} \theta \cdot \bbox{\nabla}\theta - V (\rho) \Big\}
\label{eq:2.22}
\end{equation}
where
\begin{equation}
V(\rho) = \overline{V} (\rho) + {\textstyle{1\over 8}} {(\bbox \nabla \rho)^2 \over
\rho}
\label{eq:2.23}
\end{equation}
[A similar result is obtained when a scalar field theory in $(d+1,1)$ dimensions is
reduced dimensionally in the light-cone variable ${1\over \sqrt{2}} (x^{(0)}
- x^{(d+1)})$ \cite{bib:5}.]

\subsection{d-branes} 

Following Hoppe, \cite{bib:6} we consider a Nambu-type Lagrangian for a closed extended
$d$-dimensional object moving in $(d+1,1)$-dimensional Minkowski space-tine.

\begin{equation}
L = - \int d^d\, \phi \sqrt{G}
\label{eq:2.24}
\end{equation}
$G$ is $(-)^d$ times the determinant of the induced metric
\begin{equation}
G_{\alpha\beta}= {\partial x^\mu \over \partial \phi^{\alpha}}
{\partial x _\mu \over \partial \phi^\beta}
\label{eq:2.25}
\end{equation}
where $x^\mu (\phi)$ are the coordinates of the ``$d$-brane" in space-time $(\mu =
0,1,\ldots, d+1)$, depending on parameters $\phi^ \alpha (\alpha = 0, 1, \ldots,d)$. 
[$\phi^0$ is the evolution parameter; $\phi^i(i=1,\ldots,d)$ parameterize the object at fixed
time.]  We introduce light-cone variables

\begin{eqnarray}
\tau &=& {1\over \sqrt{2}} (x^{(0)} + x^{(d+1)})
\label{eq:2.26}\\
\theta &=&  {1\over \sqrt{2}} (x^{(0)} - x^{(d+1)})
\label{eq:2.27}\\
\noalign{\hbox{and the transverse coordinates}}
x^i &\quad&(i=1,\ldots,d)
\label{eq:2.28}
\end{eqnarray}
We choose the parameterization $\tau = \phi^{0}$.   It follows that

\begin{equation}
G_{\alpha\beta} = \left( \begin{array}{rr}
G_{00}&G_{0s}\\
G_{r0}&-g_{rs}
\end{array}\right) =
\left( \begin{array}{cc} 2\partial_\tau \theta -(\partial_\tau \bbox{x})^2 &
\quad \partial _\tau \theta - \partial _\tau \bbox{x} \cdot \partial
_s \bbox{x}\\
\partial _r \theta - \partial _r \bbox{x} \cdot \partial _\tau \bbox{x} & -
\partial _r \bbox{x} \cdot \partial _s \bbox{x}
\end{array}\right)
\label{eq:2.29}
\end{equation}

\begin{mathletters}
\begin{eqnarray}
G &= & g \Gamma
\label{eq:2.30a}\\
\Gamma &\equiv & 2\partial _\tau \theta- (\partial _\tau \bbox{x})^2 + g^{rs}u _r
u_s
\label{eq:2.30b}\\
g &\equiv &\det g_{rs}
\label{eq:2.30c}\\
u_r &\equiv & \partial _\tau \bbox{x} \cdot \partial _r \bbox{x} -
\partial_r\theta
\label{eq:2.30d}
\end{eqnarray}
\end{mathletters}%
Here $\partial _\tau \equiv {\partial \over \partial \tau}
= {\partial \over \partial \phi^o}; \partial _r = {\partial \over \partial \phi^r},
 r = 1,\ldots, d;  g^{rs}$ is the inverse of  $g _{rs} \equiv \partial _r \bbox {x} \cdot
\partial _s \bbox{x}$ and will be used to move the $(r,s)$ indices. Note that the
dimensions of the $\bbox{x}$ space (indexed by $i$) and of the $\bbox {\phi}$
parameter space (indexed by $r$) are both $d$.
The Euler-Lagrange equations are conveniently presented in canonical form
\begin{eqnarray}
\partial _\tau \bbox{x} &=& -\bbox {p}/\Pi +u^r \partial_r \bbox{x}
\label{eq:2.31}\\
\partial _\tau \theta &=&  {1\over 2\Pi^2} (p^2+g) + u^r \partial _r
\theta
\label{eq:2.32}\\
\partial _\tau \bbox {p} &=&  - \partial_r ({1\over\Pi} g g^{rs} \partial _s
\bbox{x}) + \partial _r (u^r\bbox {p})
\label{eq:2.33}\\
\partial _\tau \Pi &=& \partial_r (\Pi u^r)
\label{eq:2.34}
\end{eqnarray}
with $\Pi$ and $\bbox{p}$ satisfying the constraint

\begin{equation}
\bbox{p} \cdot \partial _r \bbox{x} + \Pi \partial _r \theta = 0
\label{eq:2.35}
\end{equation}
Here $u^r$ is as given by (\ref{eq:2.30d}) [this is a consequence of  (\ref{eq:2.31}), (\ref{eq:2.35})],
and can be set to zero by appropriate choice of the parameterization.  Thereupon it follows
that
$\partial_\tau \Pi$ vanishes, so we choose $\Pi$ to be $-1$.  The equations then reduce to

\begin{eqnarray}
\partial _\tau \bbox{x} &=& \bbox{p}
\label{eq:2.36}\\
\partial _\tau  {\theta} &=& \half (p^2 + g)
\label{eq:2.37}\\
\partial _\tau \bbox{p} &=& \partial _r (g g^{rs} \partial _s \bbox {x})
\label{eq:2.38}
\end{eqnarray}
and the constraint reads
\begin{equation}
\partial _r \theta = \bbox{p} \cdot \partial _r \bbox{x}
\label{eq:2.39}
\end{equation}

The constraint is now solved by a transformation introduced by Bordemann and Hoppe.
\cite{bib:7} Rather than viewing the (dependent) variables $\bbox{p}$ and $\theta$ as
functions of the (independent) parameters $\tau$ and $\bbox {\phi}$,  a change of variables
is performed by inverting $\bbox {x} (\tau ,\bbox{\phi})$ and expressing $\bbox{\phi}$ in
terms of $\bf{\tau}$ and $\bbox{x}$, (renamed $\bbox {r}$), so that $\bbox{p}$
and $\theta$ become functions of $\tau$ (renamed $t$) and $\bbox{r}$.  It then follows
from the chain rule that the constraint (\ref{eq:2.39}) becomes

\begin{equation}
{\partial r^i\over \partial \phi^r} {\partial \theta \over \partial r^i} = p^i
{\partial r^i \over \partial \phi^r}
\label{eq:2.40}
\end{equation}
and is solved by
\begin{equation}
\bbox{p} = \bbox{\nabla} \theta
\label{eq:2.41}
\end{equation}
where the gradient is with respect to $\bbox{r}$.
Moreover, according to the implicit function theorem (see sidebar)
\begin{equation}
\partial_\tau = \partial _t + \bbox{\nabla} \theta \cdot \bbox{\nabla}
\label{eq:2.42}
\end{equation}
so that (\ref{eq:2.36}) reproduces (\ref{eq:2.41}),  since $\partial _t x^i = 0$; while
(\ref{eq:2.37}) and (\ref{eq:2.38})
 read,
respectively
\begin{equation}
\dot{\theta} + \half \bbox{\nabla} \theta \cdot \bbox {\nabla} \theta = \half g
\label{eq:2.43}
\end{equation}
\begin{equation}
\dot{\bbox{p}} + \bbox{\nabla} \theta \cdot\bbox{\nabla p}= \bbox{\nabla}(\dot{\theta}
+ \half \bbox{\nabla} \theta \cdot \bbox{\nabla}\theta) = \partial_r (gg^{rs}
\partial_s \bbox{r})
\label{eq:2.44}
\end{equation}
The over dot signifies differentiation with respect to $t$.

It remains to show that (\ref{eq:2.43}) and (\ref{eq:2.44}) are consistent with each other.  This is
achieved by recognizing that
\begin{mathletters}
\begin{equation}
gg^{rs} = {1\over(d-1)!} \epsilon^{r i _2 \ldots i _d}
\epsilon^{s j _2 \ldots j _d} g _{i _2 j _2}\cdots g_{ i _d j_d}
\label{eq:2.45a}
\end{equation}
But $g_{ij} = {\partial x^k \over\partial\phi^i} {\partial x^k \over \partial\phi^j}$, so
\begin{eqnarray}
gg^{rs} \partial _s x^i  &=& {1\over(d-1)!} \epsilon^{ri_2\ldots i_d}
{\partial x^{k_2}\over \partial \phi^{i_2}} \dots {\partial x^{k_d} \over \partial\phi^{i_d}} \epsilon^{s {j_2} \ldots j_d} {\partial x^i\over \partial \phi^s}
{\partial x^{k_2} \over \partial \phi^{j_2}}\dots {\partial x^{k_d} \over \partial
\phi^{j_d}}\nonumber\\
&=& {1\over (d-1)!} \epsilon^{ri_2\ldots i_d} {\partial x^{k_2} \over \partial
\phi^{i_2}} \cdots {\partial x^{k_d} \over \partial \phi^{i_d}} \epsilon^{ik_2\ldots k_d} \det {\partial x^k \over \partial \phi^j}
\label{eq:2.45b}
\end{eqnarray}
Note that $\det {\partial x^k \over \partial \phi _j} = g^\half$.  It now follows
that
\begin{eqnarray}
\partial _r (g g^{rs} \partial _s x^i) &=& {1\over(d-1)!} \epsilon^{r i_2 \ldots i_d} \epsilon^{i k_2 \ldots k_d} {\partial x^{k_2} \over \partial \phi^{i_2}} \cdots {\partial x^{k_d}
\over \partial \phi^{i_d}} {\partial g^\half \over \partial  x^j} {\partial x^j
\over \partial \phi^r} \nonumber\\
&=& {1 \over (d-1)!} \epsilon^{i k_2 \ldots  k_d} \epsilon^{j k_2 \dots k_d} g^\half {\partial g^\half \over \partial x^j } \nonumber\\
&=& \half {\partial g\over \partial x^i} 
\label{eq:2.45c}
\end{eqnarray}
\end{mathletters}%
Thus (\ref{eq:2.44}) implies (\ref{eq:2.43}), once an $\bbox r$-independent constant of
integration is absorbed in $\theta$.

It is seen that the equations for the ``$d$-brane"   collapse into (\ref{eq:2.43}).  But we
still need an equation for $g$, which can be obtained by differentiating $g$ with respect to
$\tau$ and using  (\ref{eq:2.42}).
\begin{equation}
 g g^{rs}
{\partial \over \partial \tau } g_{rs} \equiv \partial _\tau g = {\dot g} + \bbox{\nabla} \theta
\cdot
\bbox{\nabla} g 
\label{eq:2.46}
\end{equation}
But
\begin{eqnarray}
g^{rs} {\partial \over \partial \tau} g_{rs} &=& 2 g^{rs} \partial _r x^i \partial
_\tau \partial _s x^i \nonumber\\
&=& 2 g^{rs} \partial _r x^i \partial _s p^i \nonumber
\end{eqnarray}
Since $g^{rs} = {\partial \phi^r \over \partial x^i} {\partial \phi^s \over \partial
x^i}$, it  follows that the right side of (\ref{eq:2.46}) is  $2g {\partial \phi^s \over \partial x^i}
\partial _s p^i = 2g {\partial \over \partial x^i} p^i = 2g \nabla^2 \theta$.  Therefore if we
define
$g = 2\lambda / \rho^2$, (\ref{eq:2.46}) is equivalent to
\begin{equation}
\dot {\rho} + \bbox{\nabla} \cdot ( \rho \bbox{\nabla} \theta) = 0
\label{eq:2.47}
\end{equation}

We conclude therefore that the motion of a ``$d$-brane", moving in time on $d+1$ spatial
dimensions is governed by equations derivable from our Lagrangian (\ref{eq:2.1}), with
\begin{equation}
V (\rho) = \lambda / \rho.
\label{eq:2.48}
\end{equation}
\bigskip

\noindent{\bf Sidebar: Derivation of (\ref{eq:2.42})}\qquad
Consider a function $f$ that depends on $\tau$ and $\bbox{\phi}$: $f = f(\tau, \bbox{\phi})$.
The $\tau$ derivative differentiates $f$ with respect to the first argument.  Next express
$\bbox{\phi}$ as a function of $\tau$ and $\bbox{x}$, renamed
$\bbox{r}$.  It is clear that
\begin{equation}
\partial _\tau f (\tau, \bbox{\phi})\, = {\partial \over \partial \tau}
f(\tau,\bbox{\phi} (\tau,\bbox r)) - {\partial \over \partial \phi^s} f(\tau, \bbox{\phi}
(\tau,\bbox{r})) {\partial \phi^s  (\tau,\bbox{r}) \over \partial \tau}
\label{eq:2.49}
\end{equation}
Next observe that
\begin{equation}
d {\bbox \phi} (\tau, \bbox{r}) = {\partial \bbox{\phi} \over \partial \tau} d \tau + {\partial
\bbox{\phi} \over \partial r^i} d r^i
\label{eq:2.50}
\end{equation}
When $\bbox{\phi}$ is held constant, we have
\begin{equation}
{\partial \bbox{\phi} \over \partial \tau } = - {\partial \bbox{\phi} \over \partial r^i}
{\partial r^i \over \partial \tau}\, \biggr|_{\phi~{\rm constant}}
\label{eq:2.51}
\end{equation}
But from (\ref{eq:2.36}) it follows that the $\tau$ derivative of  $\bbox{r}$ is $\bbox{p}$,
which according to (\ref{eq:2.41}) is $\bbox {\nabla} \theta$ .  Thus (\ref{eq:2.51}) becomes
\begin{equation}
{\partial \bbox{\phi} \over \partial \tau } = - {\partial \bbox{\phi} \over \partial r^i}
{\partial \theta \over \partial r^i } 
\label{eq:2.52}
\end{equation}
Substitution in (\ref{eq:2.49}) and use of the chain rule shows that (\ref{eq:2.42}) is true,
once $\tau$ is renamed $t$.
\begin{eqnarray}
\partial_\tau f \,\Bigr|_{\tau =t} &=& {\partial \over \partial t} f + {\partial\theta\over
\partial r^i}  {\partial \phi^s \over \partial r^i} {\partial f \over \partial
\phi^s} \nonumber\\
 &=& {\partial \over \partial t} f + \bbox {\nabla} \theta \cdot \bbox {\nabla} f
\label{eq:2.53}
\end{eqnarray}

\section{Symmetries of the Model}
\subsection{Galileo Symmetries}

From the derivations, it should be obvious that $L_0$ (\ref{eq:2.2}), and also $L$
(\ref{eq:2.1}) (with obvious restriction on $V$) possess the Galileo symmetry.  Let us
record the generators of the infinitesimal transformations as integrals of the appropriate
densities; also we specify the action of the finite transformations (parameterized by
$\omega$) on the fields: $\rho
\rightarrow \rho _\omega, \theta \rightarrow \theta _\omega$, by presenting formulas for
$\rho _\omega (t,\bbox{r})$ and $\theta _\omega(t,\bbox{r})$ in terms of
$\rho(t,\bbox{r})$ and
$\theta (t,\bbox{r})$. One verifies that the generators are time independent according
to the equations of motion (\ref{eq:2.6})--(\ref{eq:2.8}), and this further implies that the
transformed fields $\rho _\omega$ and $\theta _\omega$ also solve
(\ref{eq:2.6})--(\ref{eq:2.8}), when $\rho$ and $\theta$ are solutions.

\begin{itemize}
\item{Time, space translation}

\begin{itemize}
\item{Energy}
\begin{equation}
H = \int d^d r\,{\cal E}\ , \qquad {\cal E} = \half \rho \bbox{\nabla} \theta \cdot
\bbox {\nabla} \theta + V (\rho) = \half j^2 / \rho + V (\rho)
\label{eq:3.1}
\end{equation}

\item{Momentum}
\begin{equation}
\bbox{P} = \int d^d r\, {\mbox{\boldmath$\cal P$}}\ , \qquad {\mbox{\boldmath$\cal
P$}}=
\rho
\bbox{\nabla}\theta =
{\bf j}
\label{eq:3.2}
\end{equation}
\end{itemize}

\item{Space rotation}

\begin{itemize}

\item{Angular momentum}
\begin{equation}
J^{ij} = \int d^d r\, (r^i {\cal P}^j - r^j {\cal P}^i)
\label{eq:3.3}
\end{equation}
\end{itemize}
\end{itemize}
With these space-time transformations $\rho _\omega$, $\theta _\omega$ are obtained
from
$\rho$,
$\theta$ by respectively translating the time, space arguments and by rotating the
spatial argument.\cite{bib:8}
\begin{mathletters}\label{eq:3.4}
\begin{itemize}
\item{Galileo boost}

\begin{itemize}
\item{Boost generator}
\begin{equation}
\bbox{B} = t \bbox{P} - \int  d^d r\, \bbox{r} \rho
\label{eq:3.4a}
\end{equation}
\end{itemize}
\end{itemize}
The boosted fields are
\begin{eqnarray}
\rho {\bbox{_\omega}} (t,\bbox{r}) &=& \rho (t, \bbox {r} - {\bbox{\omega}}t)\\
\label{eq:3.4b}
\theta {\bbox{_\omega}} (t, \bbox{r}) &=& \theta (t, \bbox {r} - {\bbox{\omega}}t ) +
{\bbox{\omega}}\cdot \bbox{r} - {\bbox{\omega}}^2 t/2
\label{eq:3.4c}
\end{eqnarray}
The inhomogeneous terms in $\theta _{\bbox{\omega}}$ are recognized as the well-known
Galileo 1-cocycle, compare (\ref{eq:2.21}). Also they ensure that the transformation law for
$\bf{v} = \bbox{\nabla} \theta$
\begin{equation}%
{\bf v} (t, {\bf r}) \rightarrow  {\bf v} _{\bbox{\omega}} (t, {\bf{r}}) ={\bf v} (t,
{\bf{r}} -{\bf {\omega}} t ) + {\bf{\omega}}
\label{eq:3.4d}%
\end{equation}%
is appropriate for a co-moving velocity.  Furthermore, knowledge about
the Galileo $2$-cocycle leads us to examine the $\bbox{P}$, $\bbox{B}$ bracket, and its
extension exposes another conserved generator, arising from an invariance against
translating
$\theta$ by a constant; this just reflects the phase arbitrariness in (\ref{eq:2.21}).
\end{mathletters}

\begin{itemize}
\item{Phase symmetry}
   \begin{itemize}
  \item{Charge}
\begin{mathletters}
\begin{eqnarray}
N &=& \int d^dr\, \rho
\label{eq:3.5a}\\
\rho _\omega &=& \rho
\label{eq:3.5b}\\
\theta _\omega &=& \theta - \omega
\label{eq:3.5c}
\end{eqnarray}\label{eq:3.5}
\end{mathletters}
\end{itemize}
\end{itemize}

\subsection{Connection with Poincar\'{e} Symmetry}

It is well known that a Poincar\'e group in $(d+1, 1)$ dimensions possesses the above
extended Galileo group as a subgroup. This is seen by identifying selected
light-cone components of the Poincar\'e generators ${\Bbb P}^\mu, {\Bbb M}^{\mu\nu}$ with
the Galileo generators,

\begin{eqnarray}
{\Bbb P}^\mu &=& ({\Bbb P}^-,{\Bbb P}^+, {\Bbb P}^i) \approx (H,N, P^i)
 \label{eq:3.6} \\
{\Bbb M}^{\mu\nu} &=& ({\Bbb M}^{+-}, {\Bbb M}^{-i}, {\Bbb M}^{+i}, {\Bbb M}^{ij})
\nonumber \\
{\Bbb M}^{+i} &\approx& B^i, \quad {\Bbb M}^{ij} \approx J^{ij}
\label{eq:3.7}
\end{eqnarray}
where the $\pm$ components of tensors are defined as in (\ref{eq:2.26})--(\ref{eq:2.27})

\begin{equation}
T^{(\pm)} = {1 \over\sqrt 2 }(T^{(0)} \pm T^{(d+1)}) \ .
\label{eq:3.8}
\end{equation}
(This fact is responsible for behavior in the ``infinite momentum" frame.)\cite{bib:9}
But the Lorentz generators ${\Bbb M}^{+-}$ and ${\Bbb M}^{-i}$ have no Galilean
counterparts.

A remarkable fact, first observed in Ref.~\cite{bib:7} and then again in simplified form
in Ref.~\cite{bib:5}, is that in the model (\ref{eq:2.1}) with $V(\rho) = \lambda/\rho$ [as
in the ``$d$-brane"  application, eq.~(\ref{eq:2.48})] one can define quantities that can be
set equal to the generators missing from the identification  (\ref{eq:3.7}), namely $
{\Bbb M}^{+-}$ and ${\Bbb M}^{-i}$.
This holds for arbitrary interaction strength $\lambda$; setting it to zero allows the
same construction for the free Lagrangian (\ref{eq:2.2}). We wish to determine
what role the additional generators have for the models  (\ref{eq:2.1}) and
(\ref{eq:2.2}).\cite{bib:10}

\subsection{Additional Symmetries}

We observe that the free action $I_0=\int d t L_0$, as well as the interacting one

\begin{equation}
I_\lambda = \int dt\, d^d r\, (\theta \dot{\rho} - \half \rho \bbox {\nabla} \theta
\cdot
\bbox{\nabla}  \theta - \lambda/\rho)
\label{eq:3.9}
\end{equation}
are invariant against time rescaling $t \to e^{\omega} t$, which is generated by

\begin{mathletters}\label{eq:3.10}
\begin{equation}
D = t H - \int d^d r\, \rho \theta \ .
\label{eq:3.10a}
\end{equation}
Fields  transform  according  to
\begin{equation}
\rho (t, \bbox{r}) \rightarrow \rho_\omega (t, \bbox{r}) = e^{- \omega} \rho
(e^\omega t, \bbox{r})
\label{eq:3.10b}
\end{equation}
\begin{equation}
\theta (t, \bbox{r}) \to \theta _\omega (t, \bbox{r}) =  e^{\omega} \theta
(e^\omega t, \bbox{r}) \ .
\label{eq:3.10c}
\end{equation}
\end{mathletters}%
The dilation generator $D$ is identified with ${\Bbb M}^{+-}$.   It is straight forward
to verify from (\ref{eq:2.6})--(\ref{eq:2.8}) that $D$ is indeed time independent.

More intricate is a further, obscure symmetry whose generator can be identified with
${\Bbb M}^{-i}$.  Consider the field-dependent coordinate transformations, implicitly
defined by
\begin{eqnarray}
t \rightarrow T (t,\bbox{r}) &=& t + 
{\bbox{\omega}} \cdot\bbox{r} + \half{\bbox{\omega}}^2 \theta (T,\bbox{R}) \nonumber\\
r \rightarrow \bbox{R} (t, \bbox{r}) &=& \bbox{r} + {\bbox{\omega}} \theta (T,\bbox{R})
\label{eq:3.11}
\end{eqnarray}
with Jacobian $|J|$.
\begin{equation}
J = \det
\left(
\begin{array}{rcl}
\displaystyle \frac{\partial T}{\partial t} && \displaystyle  \frac{\partial
T}{\partial r^j}
\\[2ex]
\displaystyle \frac{\partial R^i}{\partial t} && \displaystyle
 \frac{\partial R^i}{\partial r^j}
\end{array}
\right) =
\Bigl( 1 - {\pmb \omega} \cdot \bbox{\nabla} \theta (T, \bbox {R}) - \half \omega^2
\dot{\theta} (T, \bbox {R}) \Bigr)^{-1}
\label{eq:3.12}
\end{equation}
The transformation parameter ${\pmb \omega}$ has dimensions of inverse velocity.
When fields are taken to transform according to
\begin{mathletters}
\begin{eqnarray}
\rho (t, \bbox{r}) \rightarrow \rho _{\omega} (t, \bbox{r}) &=& \rho (T,
\bbox {R}) {1\over |J|}
\label{eq:3.13a}\\
\theta (t,\bbox{r}) \rightarrow  \theta _{\omega} (t, \bbox{r}) &=& \theta (T,
\bbox{R})
\label{eq:3.13b}
\end{eqnarray}\label{eq:3.13}
\end{mathletters}%
one verifies that $I _\lambda$ and $I_0$ are invariant.  This is readily seen for the
interaction term
\begin{equation}
\lambda \int d t\, d^d  r\, {1 \over \rho(t, \bbox{r})} \rightarrow \lambda \int d t\,
d^d  r\, {|J| \over \rho(T, \bbox {R})} = \lambda\int d T\,  d^d\, R  {1 \over \rho (T,
\bbox {R})}\ .
\label{eq:3.14}
\end{equation}
To establish invariance of $I_0$, it is useful to write it first as
\begin{eqnarray}
I_0 = &-&\int d t\, d^d r\,\rho (\dot{\theta} + \half \bbox{\nabla} \theta \cdot \bbox
{\nabla} \theta) \nonumber\\
&\rightarrow& - \int d t\, d^d r\, {\rho (T,\bbox{R}) \over |J|} \Bigl\{ {\partial \over
\partial t} \theta (T,\bbox{R}) + \half {\partial \over \partial  r^i} \theta (T,
\bbox{R}) {\partial \over \partial r^i} \theta (T, \bbox{R}) \Bigr\}
\label{eq:3.15}
\end{eqnarray}
The desired result follows once it is realized that the quantity in curly brackets equals
$J^2 \Big\{\dot{\theta} (T, \bbox{R}) + \half (\bbox{\nabla} \theta (T,
\bbox{R}))^2 \Big\}$.  The transformations (\ref{eq:3.13}) are generated by

\begin{eqnarray}
\bbox {G} &=& \int d^d r\, (\bbox{r} {\cal E}- \half \rho \bbox{\nabla} \theta^2)
\nonumber\\
 &=& \int d^d r\, ( \bbox{r} {\cal E} - \theta {\bbox{\cal P}})
\label{eq:3.16}
\end{eqnarray}
which is time independent according to (\ref{eq:2.6})--(\ref{eq:2.8}), and whose algebra
with the other generators show that $G^i$ can be identified with  ${\Bbb M}^{-i}$.

While we have no insight about the geometric aspects to this peculiar symmetry, the
following remarks may help achieve some transparency.

Observe that the Galileo generators can be expressed in terms of $\rho$ and $\bbox{j}$ or
$\rho$ and $\bf{v} = \bbox{j}/ \rho = \bbox{\nabla} \theta$.  Consequently, they are
also defined for velocity fields with vorticity, $(\bbox{\nabla} \times \bf{v} \not=
0)$, and provide well-known constants of motion for the (isentropic) Euler equations
[3].  However, the velocity potential $\theta$ is needed to form $D$ and $\bbox{G}$,
which therefore have a role only in vortex-free motion (with a specific potential or
no potential).

From the identification with Poincar\'{e} generators, we see from (\ref{eq:3.4}),
(\ref{eq:3.6}), (\ref{eq:3.10}) and (\ref{eq:3.16}) that $\rho$ is the ${\Bbb P}^+$ density
and that $\theta$ plays the role of
$x^-$. The latter identification is further suggested by the fact that for all the
transformations that are identified with Lorentz transformations, {\it viz}.~(\ref{eq:3.3}),
(\ref{eq:3.4}), (\ref{eq:3.10}), (\ref{eq:3.11}), (\ref{eq:3.13}) it is true that $$ 2 T \theta (T,
\bbox{R}) - R^2 = 2 t \theta _\omega (t,\bbox{r}) - r^2$$  where $T$ and $\bbox {R}$ are the
appropriately transformed coordinates.  The naturalness of this expression is appreciated
when it is recognized that $$x^\mu x_\mu = 2 x^+ x^- - x^i x^i$$

In the ``$d$-brane" development we saw that $\theta$ is indeed $x^-$, see (\ref{eq:2.27}).
Moreover there ${1\over \rho} \propto \sqrt{g}$ is the Jacobian of the inversion
transformation $(\tau, \bbox{\phi}) \rightarrow (t, \bbox{r} (t, \bbox{\phi}))$, so it is
quite natural that under the further transformation $(t, \bbox{r}) \rightarrow
\Big(T(t,\bbox{r} ), \bbox{R} (t,\bbox{r})\Big), 1/\rho$ acquires the Jacobian of that
transformation.  Presumably transformations (\ref{eq:3.11}), (\ref{eq:3.13}) reflect a residual
invariance or gauge-fixed ``$d$-brane" theory, but the reason for the specific form
(\ref{eq:3.11}) of the transformation is not apparent.

In applications to fluid mechanics our transformation generates nontrivial solutions of
Euler's equations, as we now explain.

\section{Explicit Solutions and their Transforms}
In order to gain insight into the peculiar diffeomorphism transformations (\ref{eq:3.11}),
(\ref{eq:3.13}), we consider its effect on some explicit solutions to Eqs.\
(\ref{eq:2.6})--(\ref{eq:2.8}), in the free
$(V=0)$ and interacting $(V=\lambda/\rho)$ cases.

\subsection{No Interaction,  V = 0}
\subsubsection{Particular Example}

With $V =0$, eq.~(\ref{eq:2.7}) decouples from $(2.6)$, and is solved by
\begin{mathletters}\label{eq:4.1}
\begin{equation}
\theta (t, \bbox{r}) = \half {r^2\over t}
\label{eq:4.1a}
\end{equation}
which, apart from selecting an origin in time and space and presenting a rotation and
boost invariant profile, is also invariant against time rescaling (\ref{eq:3.10}) and the
unconventional diffeomorphism (\ref{eq:3.11})--(\ref{eq:3.13}).  The fluid moves with a velocity
unaffected by boosts (\ref{eq:3.4d})
\begin{equation}
{\bf{v}} = {\bbox{r} \over t}
\label{eq:4.1b}
\end{equation}
The density is not determined, since the solution of the continuity equation (\ref{eq:2.6})
in $d$ spatial dimensions involves an arbitrary function of $t/r$, and of the angles
specifying $\bbox{r}$ that are denoted by $\hat {\bbox{r}} \equiv \bbox{r} / r$ \ .
\begin{equation}
\rho(t,\bbox {r}) = {f (t/r,\hat{\bbox{r}}) \over r^d}
\label{eq:4.1c}
\end{equation}
\end{mathletters}
With $\theta$ as in (\ref{eq:4.1a}), the coordinate transformations (\ref{eq:3.11}) takes explicit
form
\begin{eqnarray}
T (t,\bbox{r}) &=& t r^2 _{\bf{\omega}} \nonumber\\
\bbox{R} (t,\bbox{r}) &=& r\bbox{r} _{\bf{\omega}} \nonumber\\
J &=& r^2 _{\bf{\omega}} \nonumber\\
\bbox {r}_{\bbox{\omega}} &\equiv& \hat{\bbox{r}} + {\bbox{\omega} \over 2} {r\over t}
\label{eq:4.2}
\end{eqnarray}
so that, as stated, the transformed $\theta _{\bbox{\omega}} (t, \bbox{r}) = \theta (T,
\bbox{R}) = R^2/2T$ coincides with
$\theta (t,\bbox{r})$, while the transformed density retains the form (\ref{eq:4.1c}) but with
a different function of
$t/r$,
\begin{equation}
\rho_{\bbox{\omega}} (t,\bbox{r}) = \frac{f
\left({{tr_{\bbox{\omega}}}/{r}},
\hat{\bbox{r}}_{{\bf\omega}} \right)} {r^d r^{d+2}_{\bf\omega}} \ .
\label{eq:4.3}
\end{equation}
This coincides with $\rho(t,\bbox{r})$ for the special choice $f (t/r,\hat{\bbox{r}})$
$\propto (t/r)^{2+d}$ in (\ref{eq:4.1c}), which provides a density profile that is
invariant under the diffeomorphism (\ref{eq:3.11})--(\ref{eq:3.13}).

One may construct other ``free" solutions, for which $\theta$ remains unchanged under
(\ref{eq:3.11}), (\ref{eq:3.12}), (\ref{eq:3.13b}), while $\rho$ involves arbitrary functions. Also there
are free solutions that respond nontrivially to the transformations.  We do not pursue
specific solutions any further, because it is possible to give the general solution to
the free problem, which we now present.

\subsubsection{General Solution}

Since the pair of equations (\ref{eq:2.6}), (\ref{eq:2.7}) is first-order in time, to give a
general solution we need initial data at initial time $t_o$; owing to time-translation 
invariance, $t_o$ can be taken to be $0$, without loss of generality.
\begin{eqnarray}
\rho(0,\bbox {r}) &=& \rho _o (\bbox {r})
\label{eq:4.4}
\\
\theta (0,\bbox {r}) &=& \theta_o (\bbox {r})
\label{eq:4.5}
\end{eqnarray}
The general solution to the free version of  (\ref{eq:2.6}), (\ref{eq:2.7}), that is, at $ f =
0$, is given in terms of a vector-valued function of $t$ and $\bbox {r}$, $\bbox{q} (t,
\bbox{r})$, which satisfies
\begin{equation}
\bbox{q} + t  \bbox{\nabla} \theta _o (\bbox{q}) = \bbox{r}
\label{eq:4.6}
\end{equation}
The solution to (\ref{eq:2.6}), (\ref{eq:2.7}) then reads
\begin{eqnarray}
\rho(t, \bbox{r}) &=& \rho_o (\bbox{q}) \biggr| \det  {\partial q^i \over \partial r^j}
\biggr|
\label{eq:4.7}
\\
\theta(t, \bbox{r}) &=& \theta _o(\bbox{q}) + {t\over 2} \Bigl(\bbox{\nabla} \theta
(\bbox{q})\Bigr)^2
\label{eq:4.8}
\end{eqnarray}
Note that the velocity, which is the gradient of $\theta$, is just the initial velocity, evaluated
on $\bbox{q}$.

\begin{equation}
{\bf{v}} (t, \bbox{r}) \equiv \bbox{\nabla} _{\bbox{r}} \theta (t,\bbox{r}) =
\bbox{\nabla}_{\bbox{q}} \theta_o (\bbox {q}) = \bf{v}_o (\bbox {q})
\label{eq:4.9}
\end{equation}
Consequently, (\ref{eq:4.6}) may also be presented as
\begin{equation}
\bbox{q} + t \bf{v}_0(\bbox{q}) = \bbox{r}
\label{eq:4.10NEW}
\end{equation}
and one verifies that the free Euler equation, 
{\it without\/} the irrotational condition on
$\bf{v}$, $\bbox{\nabla}\times\bf{v}\neq0$,
\begin{equation}
\dot{\bf{v}} + \bf{v}\cdot\bbox{\nabla}  \bf{v} = 0
\label{eq:4.11NEW}
\end{equation}
is solved by $v_0(\bbox{q})$, where $v_0(\bbox{r})$ is the initial velocity and $\bbox{q}$
satisfies (\ref{eq:4.10NEW}). This of course is nothing but the description of freely moving
dust particles. Also one verifies that the form (\ref{eq:4.7}) for $\rho$ solves the continuity
equation, even when  $\bbox{\nabla}\times\bf{v}\neq0$.

Note that the free motion of dust particles leads to the conserved quantities
\begin{eqnarray}
J_f &=& \int d^d r\, \rho(t,\bbox{r}) f(v^{i_1} (t,\bbox{r}),\dots,v^{i_d} (t,\bbox{r}))
\nonumber\\
\dot J_f &=& 0 \label{eq:4.12NEW}
\end{eqnarray}
where $f$ is an arbitrary function of the velocity components.
This is an expression in the continuum formalism of the fact that velocities of the free theory
are constant. 
At $d=1$, there are additional conserved quantities
\begin{eqnarray}
I_f &=& \int dx\,  f(v(t,x)) \nonumber\\
\dot I_f &=& 0 \label{eq:4.13NEW}
\end{eqnarray}
for arbitrary functions $f$.

\subsection{With Interaction, \protect\boldmath$ V = \lambda/\rho$}
\subsubsection{Particular Example}
With interaction, the equations of motion remain coupled.
\begin{equation}
\dot \rho = - \bbox\nabla \cdot (\rho \bbox{\nabla} \theta)
\label{eq:4.10}
\end{equation}
\begin{equation}
\dot\theta = - \half \bbox{\nabla} \theta \cdot \bbox{\nabla} \theta + \lambda / \rho^2
\label{eq:4.11}
\end{equation}
Remarkably a solution exists that is similar to  (\ref{eq:4.1}): demanding rotational
and  time-rescaling invariance, $\ref{eq:4.10})$ leads to a second order equation in $r$ for
$\theta$, once $\rho$ has been eliminated with the help of $(\ref{eq:4.11})$.  A particular
solution of that equation then gives
\begin{eqnarray}
\theta ( t, \bbox{r}) &=&  \quad  -{ r^2 \over 2(d-1) t}
\label{eq:4.12}
\\
\rho(t,\bbox{r}) &=& \sqrt{2\lambda\over d} (d-1) {|t| \over r}
\label{eq:4.13}
\end{eqnarray}
The solution is not diffeomorphism invariant; see below.
The velocity flow is
\begin{equation}
{\bf{v}} = - {\bbox {r} \over (d-1) t}
\label{eq:4.14}
\end{equation}
while the current reads
\begin{equation}
\bbox{j} = \mp \sqrt {2\lambda \over d}  \hat{\bbox {r}}
\label{eq:4.15}
\end{equation}
where the sign is determined by the sign of $t$.  Evidently here $d$ must be greater than
$1$ and $\lambda$ must be positive.  (The latter requirement is natural, since in the fluid
mechanical application the sound speed $u$ for our model is $= \sqrt
{2\lambda} /\rho$.)

\subsubsection{General Solutions at d = 1}

It turns out that this model, along with a much more general class of models
with local interactions, is completely integrable in one dimension. \cite{bib:11} 
The integrals of motion can be expressed in the form
\begin{eqnarray}
I_f^{(\pm)} &=& \int dx \,\rho f\left( v \pm {\sqrt{2\lambda} \over \rho} \right) \nonumber\\
\dot I_f^{(\pm)} &=& 0 \label{eq:INT}
\end{eqnarray}
for arbitrary functions $f$. We obtain two `chiral' sectors involving
the velocity of the fluid plus (minus) the local velocity of sound.

The general solution can be obtained in this case through linearization.
It is known the non-linear Euler and conservation equations
can be mapped into a linear system.\cite{bib:3}  This is achieved by performing a double
Legendre transform on $\theta$, thereby replacing the independent variables $t$ and $x$,
with $v = {\partial\theta\over \partial x}$ and with the enthalpy $w = - {\partial \theta
\over
\partial t} - \half \Big( {\partial \theta \over
\partial x }\Big)^2$ \ .
\begin{eqnarray}
\Psi (v, w)  &=& \theta (t,x) - t {\partial \theta (t,x) \over \partial t}  - x {\partial \theta
(t,x)\over \partial x}\nonumber\\
&=& \theta (t,x) + t (w + \half v^2) - x v
\label{eq:4.23}
\end{eqnarray}
It is then true that
\begin{equation}
{\partial \Psi \over \partial w} = t, \quad v {\partial \Psi \over \partial w }- {\partial \Psi
\over \partial v } = x
\label{eq:4.24}
\end{equation}
An equation for $\Psi$ follows from the continuity equation for $\rho$,  when it is
remembered that for isentropic motion $\rho$ can be taken to be a function of $w$. In
our model $w=-\lambda /\rho^2$.  One finds
\begin{equation}
u^2 {\partial^2 \Psi \over \partial w^2} - {\partial^2 \Psi \over \partial v^2} + {\partial
\Psi \over \partial w} = 0 
\label{eq:4.25}
\end{equation}
where the sound speed $u=\bigl(\rho {\partial^2 V\over \partial \rho^2}\bigr)^{1/2}$ is
regarded as a function of
$w$. For our problem
$u^2 = 2\lambda/\rho^2 = - 2 w$. Note that the coupling strength $\lambda$ does not
occur in  (\ref{eq:4.25}). It reenters the formalism through
$w = -\lambda/\rho^2$,  $w \lessgtr 0$ for $\lambda \gtrless 0$.  

The general solution of (\ref{eq:4.25}) can be expressed in terms of 
two general functions, $F$ and $G$,  of one variable. It is straightforward to 
verify that the following expression solves (\ref{eq:4.25}), with $u^2 = -2w$.
\begin{eqnarray}
\Psi (v,w) &=& F(v+\sqrt{-2w}) -\sqrt{-2w} \, F' (v+\sqrt{-2w}) \nonumber\\
 &+& G(v-\sqrt{-2w}) +\sqrt{-2w} \, G' (v-\sqrt{-2w})
\label{eq:4.31}
\end{eqnarray}
Explicit solutions can be obtained from this expression
for
specific choices of $F$, $G$. 

We record the general time-rescaling invariant 
solutions.  
The time-rescaling {\it Ansatz\/} for $\theta$ and $\rho$ [$\theta\propto 1/t$,
$\rho\propto t$] lets us use 
(\ref{eq:4.11}) to express $\rho$ in terms of~$\theta$ and its spatial derivative. Then
(\ref{eq:4.10}) yields a second-order equation in $x$ for~$\theta$, which is solved by
expressions that involve
 two arbitrary
integration constants.  For $\lambda > 0$  we find
$$
\begin{array}{rclr}
\vrule width1in depth0pt height0pt \theta(t,x) &=& {\displaystyle{1\over 2k^2t}}  
\sinh^2 k x,  \quad {\displaystyle-{1\over 2k^2t}} \cosh^2 kx &\vrule width1in
depth0pt height0pt  {\rm{(4.25a), (4.25b)}}
\\[2ex]
\vrule width1in depth0pt height0pt \rho(t,x) &=& 
 {\displaystyle{\sqrt{2\lambda}\, k|t| \over \sinh^2kx}},
\qquad {\displaystyle{\sqrt{2\lambda}\, k|t|
\over \cosh^2 kx}} &\vrule width1in depth0pt height0pt  {\rm{(4.26a), (4.26b)}}
\end{array}
$$ while for $\lambda < 0$ 
$$
\begin{array}{rclr}
\vrule width1.3in depth0pt height0pt \theta (t,x)&=& {\displaystyle{1\over 2k^2t}}
\sin^2 kx, \quad {\displaystyle{1\over2k^2t}}\cos^2 kx &\vrule width1in depth0pt
height0pt {\rm{(4.27a), (4.27b)}}
\\[2ex]
\vrule width1.3in depth0pt height0pt \rho(t,x)&=& {\displaystyle{\sqrt{2|\lambda|}\,
k |t| \over
\sin^2 kx}},
\qquad {\displaystyle{\sqrt{2|\lambda|}\, k|t|
\over \cos^2 kx}} \ . &\vrule width1in depth0pt height0pt {\rm{(4.28a), (4.28b)}}
\end{array}
$$
\setcounter{equation}{28}%
Here $k$ is an arbitrary, positive integration constant; the second constant gives an
origin to $x$, and has been suppressed in the above by setting it to zero.  Note that the
current of the second solution at $\lambda > 0$ has the kink shape.
\begin{equation}
j = \mp \sqrt{2\lambda} \tanh  kx
\label{eq:4.20}
\end{equation}
This profile puts one in mind of soliton phenomena, and hints at the integrability
of Eqs.\ (\ref{eq:4.10}), (\ref{eq:4.11}) in one dimension.

In terms of the linearized formalism of (\ref{eq:4.23})--(\ref{eq:4.31}), the
$\lambda>0$ solutions (4.25a) and (4.26a) correspond to 
\begin{equation} F(z) = -G(z) = {z\over 2k} (1-\ln z)
\label{eq:4.30NEW}
\end{equation} while (4.25b) and (4.26b) have
\begin{equation} F'(z) = G(-z) = {z\over 2k} (1-\ln z)
\label{eq:4.31NEW}
\end{equation} The $\lambda<0$ solution in  (4.27) and (4.28) are the analytic
continuation of the above.

\subsubsection{Transforming the  d = 2 Solution}

We now exhibit the form of the solutions when the field-dependent, coordinate
transformation (\ref{eq:3.11})-(\ref{eq:3.13}) are performed
on (\ref{eq:4.12})-(\ref{eq:4.15}).  For simplicity we discuss only the $d =2$ (membrane)
case, and take $t> 0$.  The new coordinates are determined by the old ones by
(\ref{eq:3.11}) and (\ref{eq:4.12}).

\begin{mathletters}\label{eq:4.33}
\begin{eqnarray}
T&=& {\textstyle\frac34} t + \half \bbox{\omega} \cdot {\bf{r}} \pm {\textstyle\frac14}
\sqrt{(t+2\bbox\omega \cdot {\bf{r}})^2-2 \bbox{\omega}^2 r^2} 
\label{eq:4.33 a}\\
{\bf{R}} &=& {\bf{r}} + {\bbox{\omega} \over 2\bbox{\omega}^2}\lbrack - t-
2\bbox{\omega} \cdot {\bf{r}} \pm \sqrt{(t+2\bbox{\omega} \cdot {\bbox{r}}^2 -2
\omega^2 r^2}\rbrack
\label{eq:4.33 b}\\
{1\over J} &=& 1+ {\bbox{\omega} \cdot {\bf{R}} \over T} - {\omega^2 R^2\over 4T^2}
\label{eq:4.33 c}
\end{eqnarray}
\end{mathletters}
After $\theta$ and $\rho$ are transformed according to the rules  (\ref{eq:3.13}), it is
noticed that expressions are simplified by performing the Galileo boost ${\bf {r \rightarrow
r}} -
\bbox {\omega} t/ \omega^2$, according to the rules  (\ref{eq:3.4}).  (This precludes taking
the limit $\bbox{\omega} \rightarrow 0$.)  Also, time is rescaled according to (\ref{eq:3.10}),
with $t \rightarrow \sqrt {2} t$.  Finally, we define $\bbox{\omega} / \omega^2 = {\bf{c}}$,
which has dimension of velocity, and then the transformed profiles provide two solutions,
depending on the sign of the square root

\begin{mathletters}\label{eq:4.34}
\begin{eqnarray}
\theta _{\bf{c}} (t, {\bf{r}}) &=& \pm {\sqrt {2({\bf{c \cdot r}})^2 - c^2 r^2- c^4 t^2}}
\label{eq:4.34a}\\
\rho _{\bf{c}} (t,{\bf{r}}) &=& {\sqrt {2\lambda} \over c^2} \Bigg\lbrack {2( \bbox{c\cdot
r})^2 - c^2r^2-c^4t^2\over r^2\mp 2t {\sqrt{2({\bf {c\cdot r}})^2 -c^2r^2 -c^4t^2}}}
\Bigg\rbrack^{1/2} \ .
\label{eq:4.34b}
\end{eqnarray}
The velocity is 
\begin{equation}
{\bf{v_c}} (t, {\bf{r}}) = \pm \frac{2{\bf c} ({\bf{c\cdot r}}) - {\bf {r}} c^2}
{\sqrt{2({\bf{c\cdot r}})^2 - c^2 r^2- c^4 t^2}}
\label{eq:4.34c}
\end{equation}
and the current reads
\begin{equation}
{\bf{j}} _{\bf{c}} (t,{\bf{r}}) = \pm {\sqrt{2\lambda}} 
\frac{2{\bf{\hat{c}}}({\bf{\hat{c}\cdot r}})-{\bf{r}}}
{\lbrack r^2 \mp 2t \sqrt{2({\bf{c\cdot r}})^2 - c^2 r^2- c^4 t^2} \, \rbrack^{1/2}}\ .
\label{eq:4.34d}
\end{equation}
\end{mathletters}%
Note that $c$ may be replaced by $ic$ and $\rho_c$ by $-\rho_c$, to obtain another solution.

In the figures we exhibit the profiles of the interacting solutions.  We plot the original and
transformed densities, and the transformed currents $\bf{j} _c = \rho_c \bbox \nabla
\theta_c$, in terms of the variables ${\bf{r}} /t \, \, (t>0)$.  Without loss of generality,
$\bf{c}$ is taken along the $x$-axis, and its magnitude is incorporated in the dimensionless
ratio
${\bf{r}} /ct$.  The original density possesses a singularity at the origin; in the transformed
solutions the singularity is present only with the upper (negative) sign in the bracketed
expression of  (\ref{eq:4.34b}),  where its denominator vanishes at $r^2=({\bf{\hat{c}\cdot
r}})^2 = 2c^2t^2$.  The transformed currents exhibit a similar singularity. In the physical
region the argument of the square root must be positive, $2({\bf{\hat{c}\cdot r}})^2 -
r^2 - c^2 t^2 \geq 0$.  This requirement creates an envelope of validity for some of the
profiles.

\begin{figure}
$$\BoxedEPSF{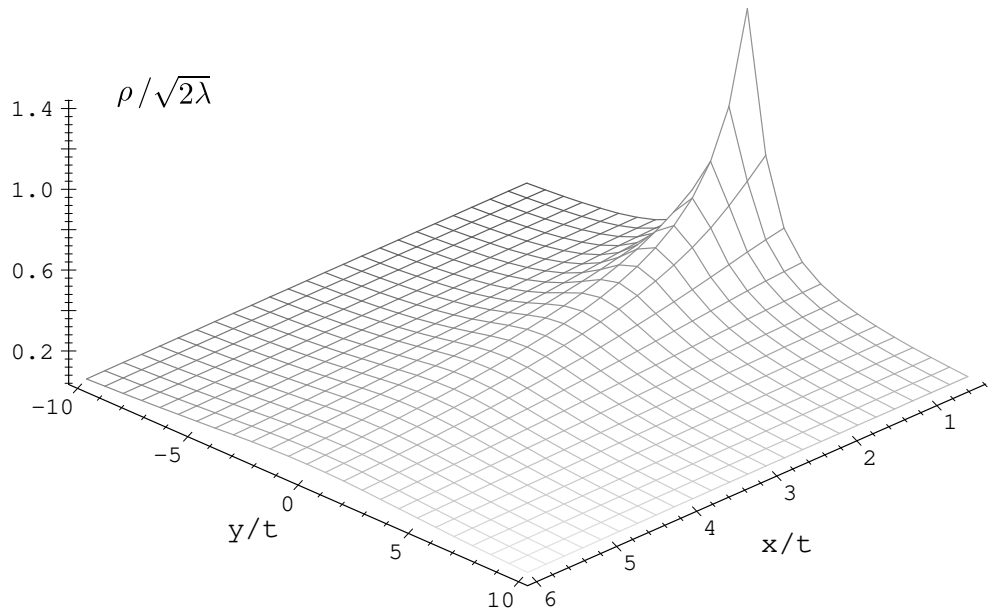 scaled 600}$$
\caption{The original density  $\rho(t,\vec r)/\protect{\sqrt{2\lambda}}$.}
\label{DBRJfig:1}
\end{figure}

\begin{figure}
$$\BoxedEPSF{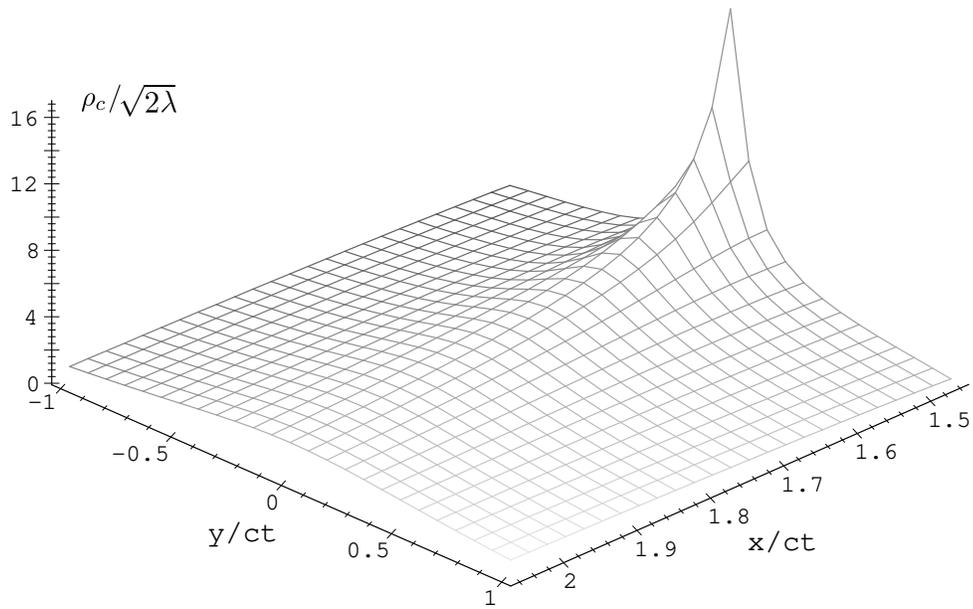 scaled 600}$$
\caption{The transformed density  $\rho_c(t,\vec r)/\protect{\sqrt{2\lambda}}$,
with the upper sign.}
\label{DBRJfig:2}
\end{figure}

\begin{figure}
$$\BoxedEPSF{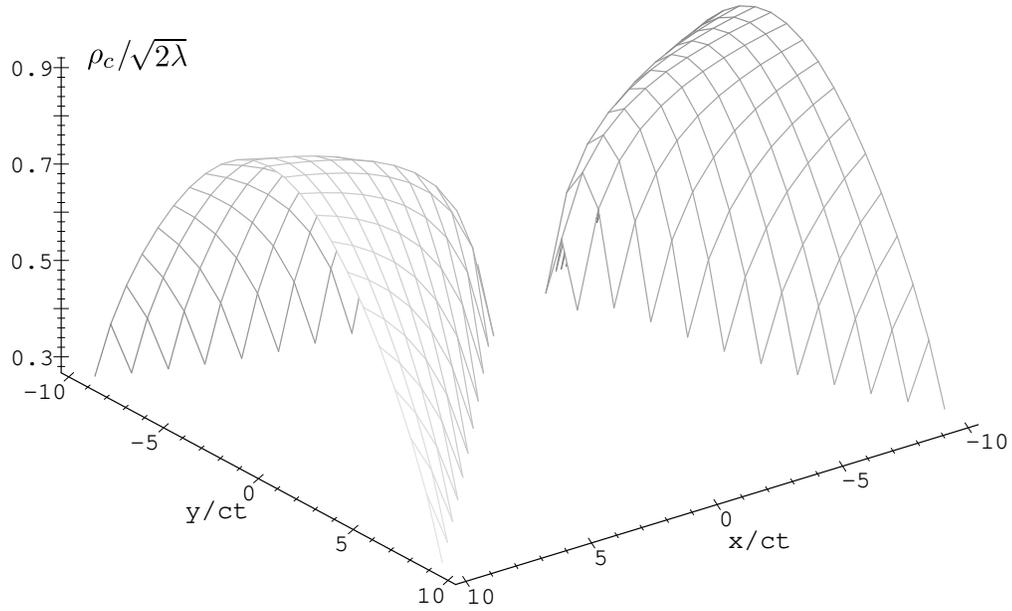 scaled 600}$$
\caption{The transformed density  $\rho_c(t,\vec r)/\protect{\sqrt{2\lambda}}$,
with the lower sign. The envelope defining the physical region
is at $x^2-y^2=c^2t^2$.}
\label{DBRJfig:3}
\end{figure}

\begin{figure}
$$\BoxedEPSF{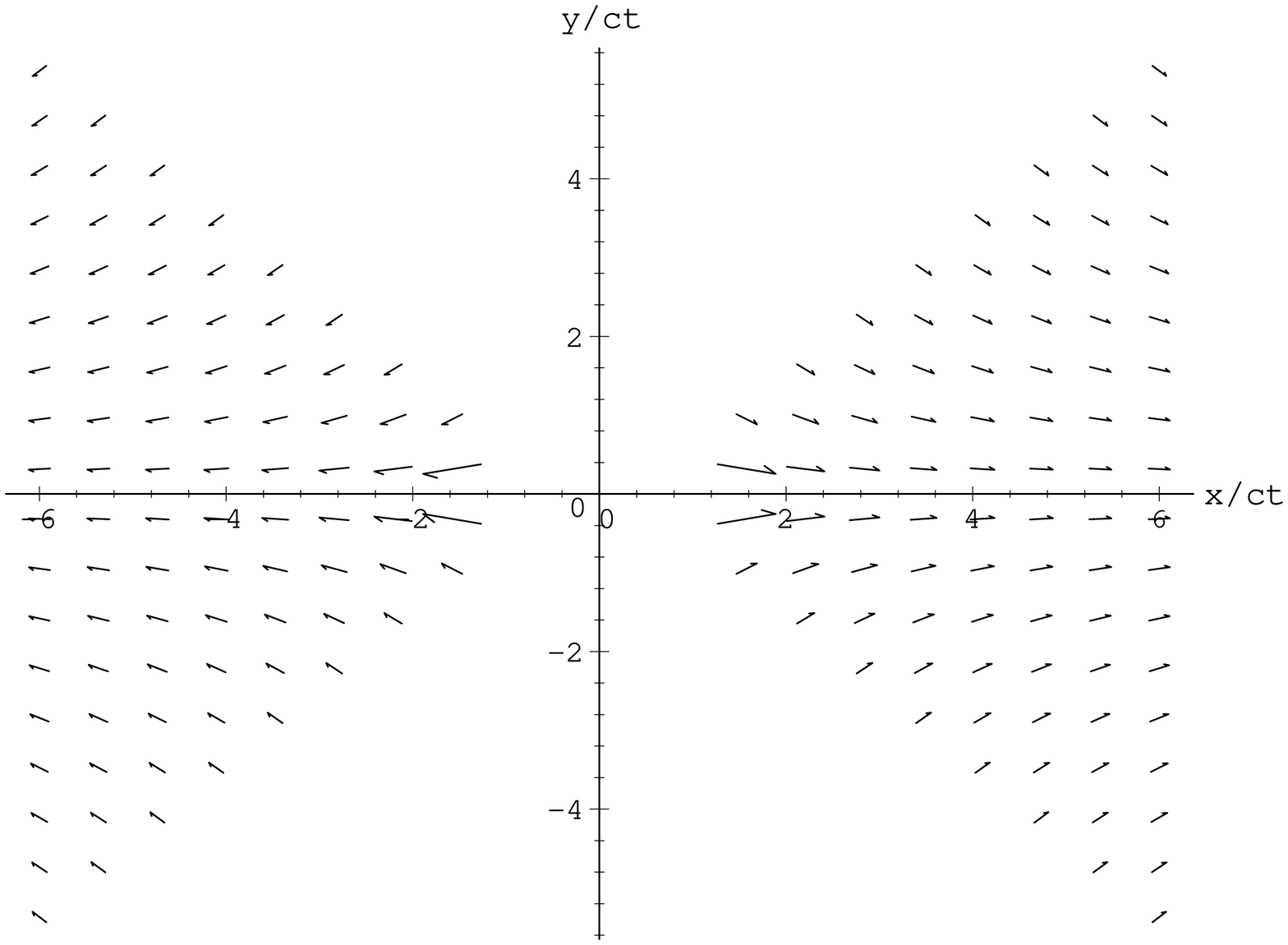 scaled 500}$$
\caption{The transformed current  $\vec j_c(t, \vec r)/\protect{\sqrt{2\lambda}}$,
with the upper signs.  The envelope defining the physical region
is at $x^2-y^2=c^2t^2$.}
\label{DBRJfig:4}
\end{figure}

\begin{figure}
$$\BoxedEPSF{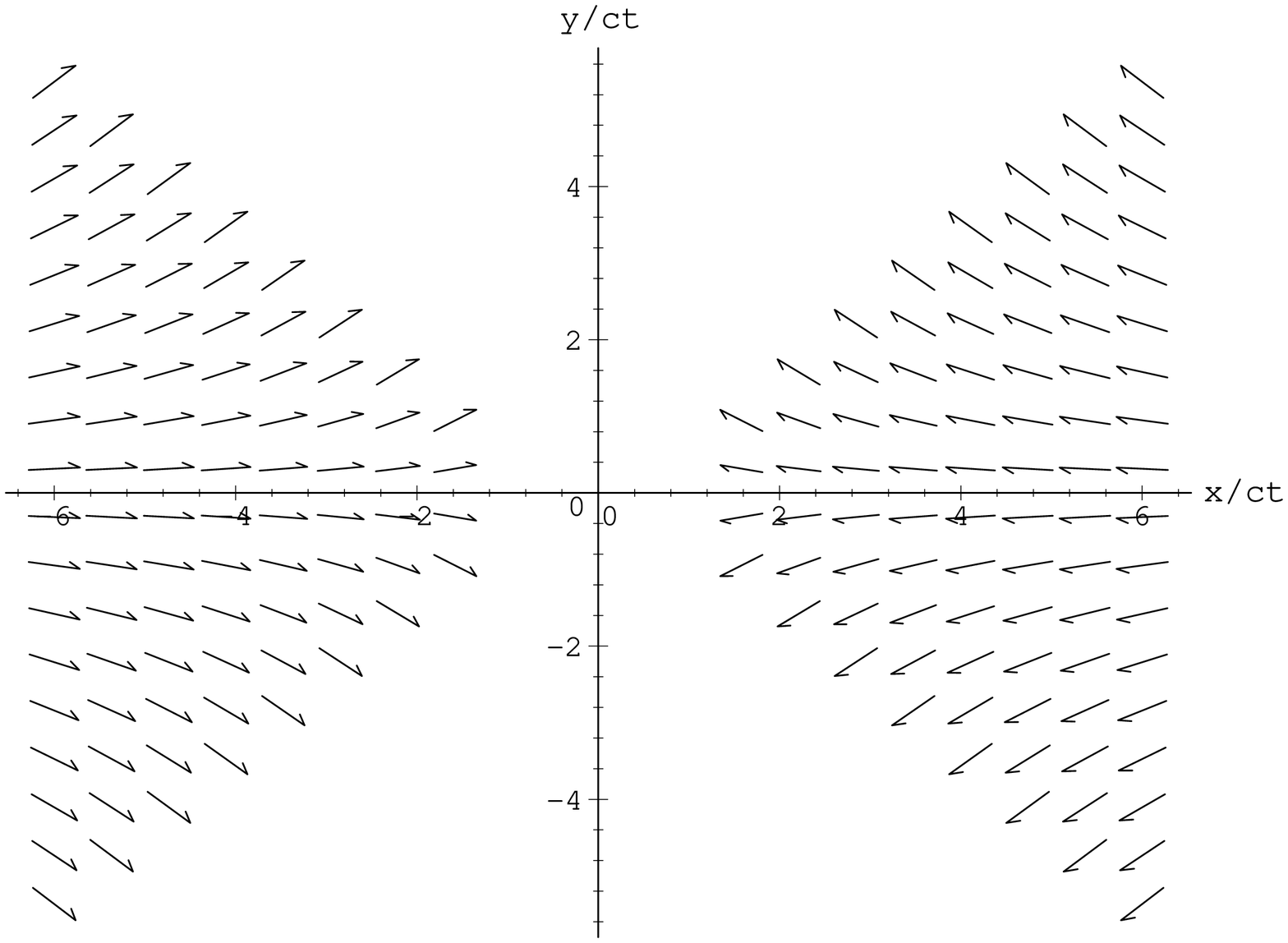 scaled 450}$$
\caption{The transformed current $\vec j_c(t,
\vec  r)/\protect{\sqrt{2\lambda}}$, with the lower signs. The envelope
defining the physical region is at $x^2-y^2=c^2t^2$.}
\label{DBRJfig:5}
\end{figure}

\end{document}